\documentclass[10pt]{iopart}
\usepackage{amsfonts}
\usepackage{bm}%
\usepackage{graphicx}
\usepackage{graphics}
\usepackage{amssymb}
\usepackage{hyperref}
\usepackage{latexsym}
\usepackage{dcolumn}% Align table columns on decimal point

\begin{document}

\title{Hydrodynamics of compressible superfluids in confined geometries} 
\author{Abdul N Malmi-Kakkada$^1$, Oriol T Valls$^1$,Chandan Dasgupta$^2$} %abdiop adjusted according to iop
%\email{malmikakkada@physics.umn.edu}
\address{$^1$School of Physics and Astronomy, University of Minnesota, 
Minneapolis, Minnesota 55455} 
\address{$^2$ Centre for Condensed Matter Theory, Department of Physics,
Indian Institute  of Science, Bangalore 560012, India}
%\author{Oriol T. Valls}
%otviop \email{otvalls@umn.edu}
%\altaffiliation{Also at Minnesota Supercomputer Institute, University of Minnesota,
%Minneapolis, Minnesota 55455}
%\affiliation{School of Physics and Astronomy, University of Minnesota, 
%Minneapolis, Minnesota 55455}

%\author{Chandan Dasgupta }
%\email{cdgupta@iisc.ernet.in}
%\altaffiliation{Also at Jawaharlal Nehru Centre %cdn 
%for Advanced Scientific Research, Bangalore 560064, India}
%\affiliation{Centre for Condensed Matter Theory, Department of Physics, 
%Indian Institute  of Science, Bangalore 560012, India}

\date{\today}

\begin{abstract}
We present a study of the hydrodynamics of  compressible
superfluids in confined geometries. We use a perturbative 
procedure in terms of the dimensionless expansion parameter
$(v/v_s)^2$  where $v$ is the typical speed of the flow and $v_s$ the
speed of sound. A zero value of this parameter corresponds to the
incompressible limit. We apply the procedure to two specific
problems: the case of a trapped superfluid with a  
gaussian profile of the local density, and  that of a superfluid
confined in a  
rotating obstructed cylinder.  We find that the corrections due to finite
compressibility which are, as expected, negligible for liquid He, are 
important but amenable to the perturbative treatment for typical 
ultracold atomic systems.
\end{abstract}

\pacs{47.37.+q, 67.25.bf, 67.85.De, 47.27.nd} %otvf

% \maketitle

\section{Introduction}

The hydrodynamics of superfluids confined in containers or channels of complex geometry is relevant
to a variety of experimentally studied systems. 
%abdiop
%Superfluids confined in porous media
%such as Vycor glass or containers packed with fine  %otvf
%powder~\cite{bill,porous2,porous3,porous4} are well-known examples of systems in which 
%the superfluid is confined in channels of complex geometry. 
%abdr removed the following inorder to deemphasize NCRI
%otvr1 good idea but I keep NCRI definition here
The absence of friction in a superfluid and the irrotational nature of superfluid flow (in the absence
of vortices) lead to a variety of unusual
hydrodynamic effects that % cdn2 NCRI defined later
%, including a decrease in
%the moment of inertia of a container filled with the superfluid.
% decreases when the fluid undergoes a transition from
%the normal state to the superfluid state. The value of this decrease 
%These effects 
depend crucially on the confining geometry.

A large number of experimental investigations into superfluidity in trapped, ultracold
atomic systems~\cite{review1,review2,review3} have been carried out 
in recent years. %abdf ref added
Various signatures of superfluidity, such as persistent flow, reduction in the moment of inertia due to the
frictionless nature of the superfluid (the so-called non-classical rotational inertia 
(NCRI)) and %cdn2 defined NCRI here
formation of quantized vortices have been observed in both 
bosonic~\cite{bec1,bec2,bec3,becv} and fermionic~\cite{fermi} systems. %otvff
In all these experiments, the superfluid is confined in a small region by an 
external trapping potential. %cdn2 emphasized confinement
While the early experiments on such systems were carried out for traps
with simple geometry, more 
recent ones have begun to explore the properties
of superfluid condensates in traps with a more complex 
structure. Superfluid flow in a toroidal trap has been observed~\cite{bec3}, 
and the effects of a repulsive optical barrier that tends to block the superflow have been investigated
in recent experiments~\cite{bec4,bec5}. %cdnew experiments on blocked ring
%and experimental conditions under which a ring-shaped
%optical trap can be realized have been suggested~\cite{ring}. 
Studies of superfluid hydrodynamics in containers with complex geometry
are obviously relevant for understanding the results of experiments
on superfluidity in atomic systems confined in such traps.

%otvr1 some edits below
Motivation for studies of superfluid hydrodynamics in confined geometries 
is also provided by
%reports~\cite{chan1,chan1a,chan2,reppy,kojima,reppyn} of NCRI in %abdiop 
reports~\cite{chan1,balibar} of an abrupt change in the resonant period of %cdn2 wording changed & no of references reduced   
%of an abrupt change in the period of torsional oscillation
a torsional oscillator filled with solid $^4{\rm He}$, initially interpreted as
NCRI in solid $^4{\rm He}$, at sufficiently low temperatures.
While the
interpretation  is controversial~\cite{chan_new}, a possible explanation
~\cite{prokofiev1,prokofiev2,dgt,us1} is that superfluidity occurs in solid $^4{\rm He}$ along
extended crystal defects such as dislocation lines and grain boundaries %abdiop
% Since such defects
which form complex disordered structures.
Studies of the flow properties of a superfluid confined in channels
and networks of irregular geometry are obviously useful
%for a critical assessment of the %abdiop
for assessing the validity of such theories.
%abdr/abdiop edited and re-organized until here

%otvr1 some changes here too
The hydrodynamics of superfluids confined in containers of simple geometries, 
such as spherical, cylindrical or rectangular, has been
studied extensively~\cite{fetter,fetter1} in the past. These studies,
all in the incompressible limit, %otv
were  recently extended~\cite{usold} by two of us
to more complex geometries, such as wedges and blocked rings, both in the case 
where  
there are no vortices (so that the superfluid
flow is irrotational) and the case where a single vortex was present. % were 
%considered in that study. 
The same authors carried
out a study\cite{us1} on the effects of superfluidity  
along grain boundaries in a two-dimensional bosonic system.
%by the two of us in~\cite{us1}.%abdr added words on relating the current study to Ref[15]
While the %otv
results of these studies %otvf
could be applied to experiments involving superfluid 
$^4$He, they %otvf
were not directly applicable
to cold atomic systems because of the incompressibility
assumption. %otvf
This assumption constrains the local
density in equilibrium to be uniform throughout the system. While this is an 
extremely  good approximation for
superfluid $^4$He, it is not a good
one for cold atomic systems in which the presence
of a confining potential causes the equilibrium density to be substantially inhomogeneous. This inhomogeneity has significant
effects on the superfluid properties of the confined atomic system, as found in both experimental~\cite{review1} and
theoretical~\cite{sheehy,fetter3} investigations. 
Therefore, it was doubtful 
if the  results of these earlier %otvf
studies~\cite{us1,usold} would be valid for %otvf %abdr added Ref[15]
cold atomic systems.
%otvff no break
For example,  calculations in Ref.~\cite{usold} showed that the %otvf
velocity field  for a superfluid confined in a two-dimensional wedge with opening angle $\beta > \pi$ 
diverges at the tip of the wedge for {\it any} nonzero value of the angular velocity $\Omega$ of the wedge about an axis perpendicular
to it and passing through its tip. %otvf
This divergence could be removed by the nucleation of a single vortex. This 
implies that either a normal region near the tip of the wedge or a vortex must be
present for any nonzero value of $\Omega$. The size of the region near the tip where the velocity exceeds the critical velocity was
estimated to be too small to be experimentally observable for liquid $^4$He, 
but it was found that it may be observable in %cdn %otvf
cold atomic systems. However, the validity of
the results for %otvff
cold atomic systems could not be established because the
calculation was carried out for an incompressible superfluid. %cdn2 added new line
%otvr1 emphasis changed below
In general, firm conclusions for cold
atomic systems cannot be drawn %otvf
from hydrodynamic
calculations  performed under the assumption of
incompressibility and uniform equilibrium density. Clearly, a method
in which this assumption is removed
is needed for studies of the hydrodynamics of these systems.   
%abdr added following sentence on Gross-Pitaevskii equation
%otvr1 moved GP to next paragraph

To accomplish this purpose, we start, in this paper, with the %otvf
hydrodynamic equations for a compressible superfluid.
Although the effect of compressibility in cold atomic systems has been
studied via the Gross-Pitaevskii (GP) equation~\cite{fetter3,pethick}, %cdnew added reference
it is simpler for our purpose
of perturbatively studying the  
compressibility corrections,
to start directly with the hydrodynamic limit   %cdnew change in wording
and the associated coarse grained equations based
on conservation of mass and momentum. 
The hydrodynamic equations we consider can be obtained~\cite{fetter3,pethick} from the GP equation if a quantum stress term, 
known~\cite{stringari} to be unimportant in the hydrodynamic regime, is neglected. %cdnew new sentence
The procedure we use to solve %abdiop 
the hydrodynamic equations is based on an expansion in  %otviop1A
parameter $(v/v_s)^2$, the square of the Mach
number, where $v$ is some characteristic speed of the
problem and $v_s$ is the speed of sound. Such
expansions have been previously used\cite{mach,mach1} in
other  quantum fluids problems. %otviop1A 
Our expansion procedure leads to linear
differential equations, which makes it much easier to find
analytic solutions: this is a considerable advantage of our method. 
At zeroth order in this small parameter
one recovers the incompressible results, since $v_s$ is then formally
infinite. Our expansion, as it will be seen, is particulary convenient
to the study of situations where a flow is imposed on the system
by external means. 
We then proceed to apply this procedure to two specific %abdr modified sentence below in order to better explain our choice of scenarios
%otvr1 edits
situations in this category. Although these have been chosen largely because analytic 
solutions in the incompressible limit 
can either be easily obtained or already exist, %otviop1A
both of these situations have been realized in experiments on cold
atomic systems. %cdnew experimental realizations
In the
first case, we assume that external constraints confine the superfluid
in such a way as to produce a gaussian profile for the local
density of the stationary superfluid, a situation
similar to that considered experimentally in Ref.~\cite{gauss1} and theoretically in Refs.~\cite{sheehy,gauss2}. %cdnew reference to expt
The specific force
fields required to establish this distribution drop out
of the equations: only the resulting density
distribution matters. We then, for this example,
assume that a flow corresponding to one quantum of axial circulation (i.e. a single vortex) %cd
is established in such a way that it is a solution of the zeroth order hydrodynamic equations, and
evaluate the first order corrections to the velocity field and to the %otvf
density due to finite compressibility. In the second problem we
consider afresh the obstructed cylinder situation previously studied\cite{usold}
in the incompressible limit and again evaluate the first order corrections
to both components of the velocity field, and to the density,
due to the finite compressibility. The geometry considered here is similar to that 
of recent experiments~\cite{bec4,bec5} on Bose-Einstein condensates in a 
toroidal trap with a repulsive barrier. %cdnew reference to expt
In both cases we find, as expected, that %abdiop
the corrections are vanishingly small for liquid He.
%, where the incompressibility
%assumption is obviously well founded. 
On the other hand, we find that for
typical cold atomic systems the corrections due to finite compressibility
are often not negligible but that %, on the other hand, 
they are sufficiently
small to be amenable to our perturbative solution.
%abdr added the following sentence
%otvr1 edited it
The observation that corrections due to finite compressibility in cold atomic systems 
in the hydrodynamic regime are amenable to  perturbative
solution for the two widely different 
problems considered here 
%appears to be a clear indication that,
is interesting because it suggests that similar perturbative treatments would be possible for other problems
of interest in studies of superfluidity in cold atomic systems.  
%under consideration, our result is of 
 %abdiop 
%this property can be 
%considered to be general. 

%cd added new text
The rest of this paper is organized as follows. In section~\ref{methods}, we present the details of
the perturbative method of calculation used here. The results obtained from application
of this method to the problems mentioned above are described in detail in 
section~\ref{result}. Section~\ref{conclusion} %otvf space %abd2 - added in before detail
contains a summary of the main results and concluding remarks.

\section{Methods}
\label{methods}

%Hydrodynamic equations
As explained above, our objective
is to study the effect of
compressibility on superfluid hydrodynamics
in confined geometries, starting with the
results for incompressible fluids. %abdr added reference to Rayleigh 
The behavior of compressible normal liquids has been studied as
early as 1883~\cite{rayleigh}.
%otviop In this section, we describe the method we use.
Starting with the general equations governing inviscid fluid flow %abdr clarified superflow
-the continuity and Euler equations - %abdr replaced N-S equation with Euler
we use a perturbative method %abdiop 
to study the effect of finite compressibility 
in the limit where the perturbation parameter, while %cdiop1
low, is nonzero. We will see that this is a realistic
limit for cold atomic systems of experimental interest. 
%This procedure, %cdnew
%otvr1 where as mentioned above 
Our perturbative method has similarities  to that 
used in Ref.~\cite{eckart}.
That study  focuses %abdiop from here to end of subsection 
on the propagation of sound waves, 
a limit  where the perturbation parameter 
cannot be assumed to be much smaller than unity.
The small dimensionless parameter associated with the 
perturbative expansion is the Mach number, the %abdiop
ratio of the characteristic fluid velocity to the  sound speed.
For quantum systems, such expansions have been 
used earlier in Ref.~\cite{mach,mach1} which address a very different problem of 
the critical speed for the nucleation of vortices in superfluid flow around 
a disk %cdn2 small changes
as compared to our 
present work on  flow patterns %and nonclassical rotational inertia
for superfluids in confined geometries. %abdiop
The procedure will be illustrated by calculating, for two examples of
confined superfluids, %abdiop
the corrections to the velocity field and 
the density distribution in the low temperature limit
where viscosity effects can be neglected. %abdr added low-T limit
%abdr added refernce to Eckart paper below
%otvr1

\subsection{General}

As a simpler alternative to deriving the equations of compressible %abdiop 
superfluid hydrodynamics via the GP equation, we start 
with the fundamental hydrodynamic equations governing
fluid flow in the steady state i.e. 
mass conservation as given by the continuity equation: %otvr1
\begin{equation}
\label{cont}
\nabla\cdot(\rho \vec{v}) = (\nabla\rho)\cdot\vec{v} + \rho(\nabla\cdot\vec{v}) = 0
\end{equation}
and momentum conservation %otvr1
as given by the Euler equation: %abdiop1 deleted reference to Navier-Stokes
%(dissipationless limit
%of the Navier-Stokes equation): %abdr replaced N-S equation with Euler
\begin{equation}
%\frac{\partial \vec{v}}{\partial t} + 
(\vec{v}\cdot\nabla)\vec{v} = -\frac{\nabla p}{\rho} + \frac{\vec{f}}{\rho} %cdn
\label{ns}
\end{equation}
where $\rho$ is the mass density, 
$\vec{v}$ represents the velocity field,
$p$ the pressure, and $\vec{f}$ is %otvf
the external force per unit volume. %otviop1D
The steady state assumption means, as usual, that we are averaging over
microscopic scale time fluctuations. 
Hydrodynamics %abdiop
can also be derived by
starting from microscopic or
quasi microscopic equations of motion and then coarse graining. When one does  %cdnew few edits
that from the GP equations\cite{fetter3,pethick,mach,mach1} 
one obtains in the Euler
equation an additional quantum stress term. This term need not be included here
for two reasons: first, as explained on page 170 of Ref.~\cite{pethick} 
(see also Ref.~\cite{stringari})
the order of magnitude of this term
(which involves third derivatives of the density)
is down by a
factor of $(\ell/L)^2$ where $\ell$ is a microscopic quantum length and $L$ the
characteristic  length associated with macroscopic pressure variations:
it is hence negligible for the hydrodynamic problems considered here. Secondly,
this term (involving as it does density derivatives) vanishes at $\kappa=0$ and,
since it appears in the same way as the term $\vec{f}$, it would similarly acquire
an explicit  factor of $\kappa$ in Eq.~(\ref{ns1a}) below. It is therefore a second
or higher order correction in our small parameter.
%abdiop1 added the following lines 
%otviop1B edited
Short range fluctuations may exist, %cdiop1
just as they do in classical fluids, 
but
they are averaged over the macroscopic distance $L$ in the hydrodynamic limit. 
Such a scale clearly exists in experimental Bose systems:  
the Thomas-Fermi
radius of trapped Bose gases can be between two or three orders of magnitude 
larger than the microscopic coherence length. From experiments on
superfluid flow in Bose systems \cite{bec1,lambda,lambda1} it can be seen that the characteristic
Mach number squared, our dimensionless expansion parameter %abdiop2 
$(v/v_s)^2$, is  of order $10^{-2} - 10^{-4}$ and hence small in these systems.
The Thomas-Fermi approximation predicts an abrupt drop to zero
in the density of the condensate beyond the Thomas-Fermi radius, but
this is an artifact of the Thomas-Fermi method:
the actual variation in density is smoother. %abdiop2 added TF clarification
%otviop2 above edited
This does not affect the validity of our approach
just as the abrupt density drop near a wall does not invalidate
classical hydrodynamics given that a large region over which  
the parameter $(v/v_s)^2$ is small exists. %abdiop2 clarification & minor correction 
%abdiop1 addition ends here

We consider $\rho$ to be a function of %abdr introducing barotropic 
$p$ only (barotropic limit).  %otvr1 abdiop
%abdiop1 added reference/clarification of barotropic
%otviop1F edited it
This limit applies\cite{foot}
in the very low temperature case that we study: at zero temperature $p$
can only be 
a function of $\rho$. %otviop2
%abdiop1 end barotropic clarification here
Using the definition of compressibility ($\kappa = \frac{1}{\rho} \frac{\partial \rho}{\partial p}$),
%\begin{equation}
%\kappa = \frac{1}{\rho} \frac{\partial \rho}{\partial p},
%\end{equation}
Eq.(\ref{ns}) becomes:
\begin{equation}
\label{ns1a}
\rho^2\kappa(\vec{v}\cdot\nabla)\vec{v} = -\nabla\rho + \rho\kappa\vec{f}.
\end{equation}
%otvr1 above This is also referred to as the barotropic Euler equation. %abdr finished introducing barotropic
We now start the  perturbative calculation by writing:
\begin{equation} %otviop2
\label{zeroone}
%\begin{eqnarray}
\rho = \rho_0 + \rho_1
\end{equation}
\begin{equation}
\label{zeroonev}
\vec{v} = \vec{v}_0 + \vec{v}_1
\end{equation}
%\end{equation} %otviop
where the zero index denotes quantities in the 
incompressible ($\kappa = 0$) limit
%corresponding to $\nabla \rho_0 = 0$ and $\nabla.\vec{v}_0 = 0$
and the index one in $v_1,\rho_1$ denotes  %cdn
the changes
in velocity field and density distribution due to
the finite
compressibility. 
Substituting Eqs.~(\ref{zeroone}) and (\ref{zeroonev})
into Eqs~(\ref{cont})  and (\ref{ns}), 
the Euler 
equation at zeroth order takes the form:%abd2 fixed grammar
\begin{equation}
\label{ns0}
\nabla\rho_0= \rho_0\kappa\vec{f}
\end{equation}
which reflects the fact that at $\kappa=0$ it would
take an infinite force to induce a density gradient (here and below, we consider the
product $\kappa f$ to be finite, of order unity). %cd added a line of text
%otviop1 I moved abdul's stuff to the end of sct 2
For future convenience, let us assume that such a density gradient 
has somehow been induced by external means. In that
case the zero order continuity equation would take the
form:
\begin{equation}
\label{cont0}
(\nabla\rho_0)\cdot\vec{v}_0 + \rho_0(\nabla\cdot\vec{v}_0) \equiv \mathcal{D}^0
%otvf dots changed to \cdot here and everwhere later
\vec{v}_0  = 0,
\end{equation}
which reduces to the usual form $\nabla\cdot\vec{v}_0=0$
in the absence of external forces. Here 
we have introduced the operator:
\begin{equation}
\label{dzero}
\mathcal{D}^0 \equiv \rho_0(\nabla)\cdot + (\nabla\rho_0)\cdot.
\end{equation}
Proceeding now to  first order,  the corresponding terms
in the continuity equation yield:
\begin{equation}
\label{cont1}
-(\nabla\rho_1)\cdot\vec{v}_0 = \mathcal{D}^0 \vec{v}_1 + \rho_1(\nabla\cdot\vec{v}_0) 
\end{equation}
while from those in the Euler equation we have:
\begin{equation}
\label{ns1}
\rho_0^2\kappa(\vec{v}_0\cdot\nabla)\vec{v}_0 = -\nabla\rho_1 + \rho_1\kappa\vec{f}.
\end{equation}
Taking the scalar product of Eq.~(\ref{ns1}) with $\vec{v}_0$
%one obtains: %abdiop
%\begin{equation}
%-(\nabla\rho_1)\cdot\vec{v}_0 = \rho_0^2\kappa\vec{v}_0\cdot(\vec{v}_0\cdot\nabla)\vec{v}_0 - 
%\rho_1\kappa\vec{f}\cdot\vec{v}_0,
%\end{equation}
%and one can then
and making use of Eqs.~(\ref{cont1}) and (\ref{cont0}) one obtains: %cd
\begin{equation}
\label{ns11}
\mathcal{D}^0\vec{v}_1 = \rho_0^2\kappa\vec{v}_0\cdot(\vec{v}_0\cdot\nabla)\vec{v}_0.
\end{equation} 
By this procedure the external force has been eliminated from the equations.
The reason this is possible is that the only role of the force is
to impose the zeroth order density profile, $\rho_0$, which alone has
physical meaning.
Eqns.~ (\ref{cont0}), (\ref{cont1}) and (\ref{ns11}) are the basic set of
equations needed. In general, the best course to obtain the
first order results, after getting the zeroth order solution,
is to solve first Eq.~(\ref{ns11}) and then obtain the first order density
profile from Eq.~(\ref{cont1}).
%Based on these equations, 
%we will look at two different
%scenarios - one in which an external force is
%imposed on the fluid i.e. ($\vec{f}\neq0$) and 
%another one, corresponding to one
%of the confined geometries considered in
%Ref.~\cite{usold}, in which no external force is applied. %cdn2 deleted a few lines to avoid repetition

We now verify the physical
meaning of the dimensionless  perturbation
parameter associated with the low compressibility
limit, as  discussed above.
Dividing through Eq.~(\ref{ns1}) by $\rho_0$, introducing the %cdn
average speed of sound $v_s$ via $\rho_0\kappa = {v_s^{-2}}$
and introducing the dimensionless variable
$\widetilde{\rho} \equiv {\rho_1}/{\rho_0}$
(i.e. the dimensionless density correction) we have:
\begin{equation}
\label{nodim}
-\nabla\widetilde{\rho_1} + \widetilde{\rho_1}\kappa\vec{f}= 
(\frac{\vec{v}_0}{v_s}\cdot\nabla)\frac{\vec{v}_0}{v_s},
\end{equation}
where we recall that the product $\kappa f$ %cd
must be viewed as finite. We see from
this result that the dimensionless perturbation
parameter associated with the correction to the density
due to compressibility is indeed of order $({v_0}/{v_s})^2$.
Proceeding in a similar way to evaluate the order of 
magnitude  of the correction to the velocity ($\vec{v}_1$) %otvf
due to the finite compressibility, %abdiop commenting out 
it can be seen 
that the dimensionless perturbation parameter
associated with the correction to velocity field ($\vec{v}_1$)
is $\lambda \equiv ({v}_0/{v_s})^2$. %cd
%This is then our  dimensionless %abdiop 
%perturbation parameter, denoted  as
%$\lambda$.  

\subsection{Zero applied force}
%cd changes below
As explained above, the equations obtained from the perturbative analysis are quite
general and can be used when  
$\rho_0(r)$ is uniform, as well as when  
it takes on
a specific inhomogeneous form due to the application of
some external force $\vec{f}$ which need
not be specified.
%any specific zeroth order density profile for the fluid
%can be considered depending on the 
%nature of the external force applied.
%%In that general case, the equations above can be used as given.
%cdn2 removed a line to avoid repetition
In the case  where
no external force is imposed on the fluid,  the
zeroth order density distribution is of course a
constant. This applies to the calculations performed
in Ref.~\cite{usold}, for 
an obstructed annular cylinder
as explained in the Introduction. In this case it is more
convenient to start by simplifying the basic equations from
the beginning.  Since $\nabla\rho_0 = 0$ and $\nabla\cdot\vec{v}_0 = 0$, %otvf
one has for the continuity equation at first order: 
\begin{equation}
\label{cont1u}
(\nabla\rho_1)\cdot\vec{v}_0 + \rho_0(\nabla\cdot\vec{v}_1) = 0.
\end{equation}
In this limit the first order Euler equation Eq.~(\ref{ns1})
 is:
\begin{equation}
-\nabla\rho_1 = \rho_0^2\kappa(\vec{v}_0\cdot\nabla)\vec{v}_0
\label{rho1}
\end{equation}
Combining the two equations above, we obtain the
following equation for $\vec{v}_1$:
\begin{equation}
\label{ns1u}
\nabla\cdot\vec{v}_1 = \rho_0\kappa\vec{v}_0\cdot(\vec{v}_0\cdot\nabla)\vec{v}_0
\end{equation}
with the right side 
known from the solution of %otvr1 
the zeroth order equations.
%\begin{equation}
%\label{vzerou}
%\nabla\cdot\vec{v}_0 = 0
%\end{equation}
%and $\nabla \times \vec{v}_0 = 0$. %abdr added irrotational flow
%otvr1 
Since one of these equations is $\nabla\cdot\vec{v}_0 = 0$,
similar Green function methods
%otvr1 for the Laplace operator 
lead to solutions for both %abd2 minor grammar
$\vec{v}_0$ and $\vec{v}_1$.
Specializing to the curl free case (absence of vortices)
we introduce a scalar potential, $V(\vec{r})$,
such that $\vec{v}_1(\vec{r})=\nabla V(\vec{r})$. Eq.~(\ref{ns1u}) then becomes:
\begin{equation}
\label{potential}
\nabla^2V = \rho_0\kappa\vec{v}_0\cdot(\vec{v}_0\cdot\nabla)\vec{v}_0,
\end{equation}
which we solve by finding the appropriate
Green function with  specified boundary conditions.
Once this is done we solve
for $\vec{v}_1$ from:
\begin{equation}
V(\vec{r}) = \int d\vec{r}^\prime
G(\vec{r}^\prime,\vec{r})\rho_0\kappa\vec{v}_0\cdot(\vec{v}_0\cdot\nabla)\vec{v}_0
\label{po2}
\end{equation}
recalling that $\vec{v}_1=\nabla V(\vec{r})$.%abd2 changed vec with nabla
The correction to the density profile due to %otvr
nonzero compressibility - $\rho_1$ -  is calculated from Eq.~(\ref{rho1}) %cd
using the result for $\vec{v}_0$ obtained from solving the 
equation $\nabla\cdot\vec{v}_0 = 0$. %otvr1 obtained
%otvr1 from Eq.~(\ref{vzerou}).
%abdiop1 added lines here
%otviop1C edited and moved

We conclude this general discussion with a brief discussion of the
relation of our method to the Thomas-Fermi approximation.
Eqn.~(\ref{ns0}) is an
equilibrium relation ($v=0$) and in that respect
it is analogous to neglecting the kinetic 
terms in the GP equation: in this sense it resembles the 
Thomas-Fermi approximation in equilibrium. 
Corrections to the equilibrium Thomas-Fermi approximation have 
been studied previously.
For example, Ref.~\cite{ueda} keeps terms  
linear in the velocity in the context of studying collective %abdiop2 minor removed the before velocity
modes. This is  different from our %otviop2
current study where, as explained in the Introduction,
we consider induced flows. 
%abdiop1 end addition

\section{Results}
\label{result}
In this section we present the 
results
of calculations of the %otvf 
corrections to the velocity field %otvf
($\vec{v}_1$)
and density ($\rho_1$)  
due to finite compressibility 
using the perturbative method 
described in the section above. %cdn2 minor changes in wording
We study the effect of 
nonzero compressibility on the %otvf
velocity field and density distribution
of confined superfluids in two different situations.
In the first case, an external force
imposed on the fluid leads to an equilibrium
gaussian density profile. The other case, in the
$f=0$ limit, deals with the compressibility corrections for superfluid flow in an obstructed
cylinder, as discussed in Ref.~\cite{usold}.  %cdn2 minor changes in wording
%abdiop1 emphasize difference with T-F 
%otviop1C
In the case of an obstructed cylinder, the fluid
is driven by an obstruction and therefore the Thomas-Fermi
approximation is not valid as the kinetic energy
term cannot be neglected.
%abdiop1 addition ends here
\subsection{Gaussian density profile}
\label{gaussian}

We consider first the case of a cylindrical
sample where 
an external
force imposed on the 
fluid in equilibrium results in a gaussian
density profile i.e.
\begin{equation}
\label{prof}
\rho_0(r) = \rho_a e^{-(\frac{r}{\sigma})^2}
\end{equation}
where $\sigma$ is the characteristic
length scale associated with the density
profile and $r$ the cylindrical radial
coordinate. We assume the cylinder is long enough so 
that we can neglect edge effects and solve the problem
as a quasi two-dimensional one.
Density profiles other
than a gaussian can also be
considered by the same method: here we focus on 
this case as an example. 
In order to calculate
$\rho_1$ and $\vec{v}_1$ we use
equations (\ref{cont1}) and (\ref{ns11}) respectively,
which requires us to calculate first an appropriate %changed as to us  
zeroth order velocity 
field $\vec{v}_0$ corresponding to the %cd small change
incompressible limit. 

To calculate $\vec{v}_0$, we use the
zeroth order continuity equation
(Eq.~(\ref{cont0}))
with the gaussian 
density profile, $\rho_0$, specified
above. 
Calculating $\nabla\rho_0$ and
defining a velocity potential
such that 
$\vec{v}_0 = \nabla V_0(r) $,
Eq.~(\ref{cont0})
takes the form:
\begin{equation}
\label{vp}
\nabla^2 V_0 = \frac{2r}{\sigma^2} \frac{\partial V_0}{\partial r}.
\end{equation}
This equation has a variety of solutions reflecting
the many possibilities for the velocity field.
For our example, we restrict ourselves
to begin with to the case where $\vec{v}_0$ has no azimuthal dependence. 
Assuming then a purely radial solution i.e.
$V_0(r,\phi)=F(r)$ %otv \Phi(\phi)$ with 
%ith the boundary condition
%that $\Phi(\phi) = \Phi(\phi + 2\pi)$
we obtain $\mathcal{D}^0_r F(r)=0$
%the  equation:
%\begin{equation}
%\label{radpot}
%\mathcal{D}^0_r F(r)=0
%\end{equation}
where
\begin{equation}
\label{Dr0}
\mathcal{D}^0_{r} \equiv r \frac{\partial}{\partial r} + 
r^2\frac{\partial^2}{\partial r^2} -\frac{r^3}{\sigma^2}\frac{\partial}{\partial
r}.
\end{equation}
The solution is:
\begin{equation}
F(r)=C_1Ei((r/\sigma)^2)
\end{equation}
where $Ei$ is the usual exponential integral function. We obtain from this
a purely radial part
of $v_0$ 
namely:
\begin{equation}
\label{v0r}
\vec{v}_{0r} = \frac{C_1}{r} e^{(\frac{r}{\sigma})^2} \hat{r}
\end{equation}
where the integration constant
$C_1$ is a quantity with units of circulation. %cdnew correction 
%associated with the 
%radial component of $\vec{v}_0$. 
The increase in the velocity with $r$ appears strange until one recalls that %otvf
the density decreases (see Eq.(\ref{prof})) exponentially so that the %otvf
current decreases with $r$.
To this particular solution we 
add an azimuthal component corresponding to a vortex centered at
the origin. This amounts to adding to the velocity potential a term proportional
to $\phi$, which of course also satisfies the linear Eq.~(\ref{vp}). This %cdn
leads to an azimuthal component:
\begin{equation}
\label{v0p}
\vec{v}_{0\phi}=\frac{C_2}{r}\hat{\phi}
\end{equation}
Introducing this vortex in the fluid
implies that the curl of the zeroth order velocity field,
$\nabla \times \vec{v}_0 = 2\pi\delta(\vec{r}) \hat{z}$
is non-zero. 
This does not contradict the 
definition of $\vec{v}_0$ 
as a gradient of a velocity 
potential since one is dealing with a singular field.
%abdiop1 clarifying cutoff here  
%otviop1E edited
%The singularity at the origin requires the imposition of a small r cutoff 
%which will be explicitly introduced later. %cdn2 some rearrangement of text and changes in wording
The singularity at the origin requires the imposition of a
small $r$ cutoff.  
At the vortex core radius in a Bose fluid, 
the characteristic velocity of fluid flow
is of the same order of magnitude as the sound velocity \cite{fetter3}. 
The region near the %abdiop2 added citation - Fetter RMP cited before 
vortex core is also characterized by faster
density variations, making the quantum pressure term 
important. Therefore, we introduce a short distance cutoff near the vortex
of order few times the vortex core radius. Outside this
region our assumptions 
of small Mach number squared and negligible quantum pressure 
term %abdiop2 Mach number squared
hold true. 
%abdiop1 end cutoff clarification here
The circulation around the origin 
%associated with the radial field is zero and that
due to the azimuthal component of the velocity is $2\pi C_2$.
Physically, we know that the circulation must be quantized. It
follows that, as opposed to $C_1$, $C_2$ is not a completely arbitrary
constant, but must be an integer number of circulation quanta %deleted we
$h/m$. 
%otvr1 edited below
%We will assume for our numerical
%example that we have one quantum 
%of circulation. %abdr modified explanation of n=1 in the following sentences
%thus ensuring the absence of vortices
%in regions other than a small radial cut-off about
%the origin. %abdr modified explanation of n=1 %cdn2 deleted a few lines and changed wording
To make an estimate  of %otvff  
the order of magnitude of the effect of nonzero compressibility, we assume in 
our numerical
work that there is one quantum %otvff
of circulation.

We now proceed to the first order
calculation taking as our zeroth order results
the density distribution Eq.~(\ref{prof}) and
the velocity field given by the sum of Eqs.~(\ref{v0r}) and (\ref{v0p}).
%otvf rewritten
In the results presented in this subsection, we  measure lengths
in units of $\sigma$, %otvr1 edited below
and choose for our illustrative example 
%otvr1 $C_2$ corresponding to one quantum and 
$C_1=C_2$. %otvf 
%otvr1 even though a radial flow component does exist. 
This choice of $C_2$ %otvr1
%abdr clarification of $C_1=C_2$ 
leads to some formal simplifications. 
%A vector plot of the 
%zeroth order velocity field under these conditions
%is provided in Figure \ref{fig1}. %otvf no parenthesis (also later) %cdn2 fig1 has been deleted
%There, the length is given in units of $\sigma$, while the circulation
%is given in units of the flux quantum. In the plot, we have assumed that

To evaluate the first order corrections, 
we now turn to Eq. (\ref{ns11}) 
in order to solve for the correction
to the velocity field ($\vec{v}_1$). %otvf
%due to non-zero compressibility
%($\kappa \neq 0$).
We  use the Green function method
for this purpose.
It is not hard to see that the azimuthal
part of $\vec{v}_0$ does not contribute
to the right side of Eq.~(\ref{ns11}).
Hence it is sufficient to find the 
Green function corresponding
to the operator $\mathcal{D}_r^{0}$
introduced in Eq.~(\ref{Dr0}):
\begin{equation}
\label{g1}
\mathcal{D}^0_r G(r,r') = \frac{1}{r}\delta(r-r')
\end{equation}
Solving the second order
differential Eq.~(\ref{g1}),
we find this Green function to %cd
be:
\begin{equation}
G(r,r') = \frac{Ei(\frac{r_>^2}{\sigma^2})}{2}
\end{equation}
where $r_>$ is the larger
one of the two radial coordinates $r$, $r'$. 
Introducing then a velocity potential $V_1$ such that
$\vec{v}_1 = \nabla V_1(r)$, Eq.~(\ref{ns11}) leads us to
%\begin{equation}
%\label{nsp}
%\mathcal{D}^0 V_1 = \rho_0^2\kappa\vec{v}_0.(\vec{v}_0.\nabla)\vec{v}_0
%\end{equation}
$V_1$:
\begin{equation}
\label{po1}
V_1 = \int r'dr'G(r,r')[\rho_0^2\kappa\vec{v}_0\cdot(\vec{v}_0\cdot\nabla)\vec{v}_0]
\end{equation}
Evaluating the expression %otvf
on the right side of Eq.(\ref{po1})
with the Green function 
given above and the zeroth order velocity fields, we obtain %otvf
analytically the potential associated
with the correction to the velocity field %otvf
($\vec{v}_1$) due to 
nonzero compressibility. 
The quantity $\vec{v}_1$ itself can then 
be evaluated. 
%otviop1E
As already anticipated in our zeroth order results, %corrected mentioned
the divergence of the azimuthal field at $r=0$ requires
the introduction of a small $r$ cutoff, which we denote
as $b$. It is chosen as discussed above, which
also ensures %finaloa physically, it may be defined as 
%the value of $r$ at which 
that $v_0$ does not exceed the
critical velocity. The result for  $\vec{v}_1$ is found to be:
%otviop \begin{widetext}
\begin{eqnarray}
\label{v1gauss}
%\begin{eqnarray}
%otviop\label{v1gauss}
\vec{v}_1=&& \rho_a\kappa
\frac{e^{s^2}}{r\sigma^2}[\frac{C_1^3}{2}(F_1(s^2,\delta^2)) - 
\frac{C_1C_2^2}{2}(F_1(-s^2,-\delta^2) + \nonumber \\
&&2(Ei(-\delta^2) - Ei(-s^2)))]\widehat{r}  %cdn inserted )
%Ei(\frac{-b^2}{\sigma^2})}{\sigma^2} - 
% \frac{Ei(\frac{-r^2}{\sigma^2})}{\sigma^2})]
\end{eqnarray}
%\end{equation}
%otviop \end{widetext}
where we have introduced the dimensionless length $s\equiv r/\sigma$
and the dimensionless cutoff
$\delta \equiv {b}/{\sigma}$.
We have also introduced the auxiliary function $F_1$ as:
\begin{equation}
F_1(s^2,\delta^2) = -\frac{e^{\delta^2}}{\delta^2} + \frac{e^{s^2}}{s^2} -%abd2 changed arguments of F_1 
Ei(\delta^2) + Ei(s^2).
\end{equation} %cdquestion: should the arguments of F_1 in the above equation be s^2, \delta^2?

Although the zeroth order velocity has both radial and azimuthal
components, the correction is purely radial. Nevertheless, the azimuthal
component of $\vec{v}_0$ has a large effect on the result via the 
$C_2^2$ dependence of the second %otvf
term in Eq.~(\ref{v1gauss}).  
When $C_1=C_2$ it is possible to further simplify
the result. Expressing in that case $\vec{v}_1$ in 
the natural dimensionless form, that is, 
in units of $C_1 /\sigma \equiv v$ we have:
%\begin{widetext}
\begin{equation}
\label{v1gauss2}
\frac{\vec{v}_1}{v} = \frac{1}{2}(\frac{v}{v_s})^2 \frac{e^{s^2}}{s} %abd2 added \hat{r}
(H(\delta) - H(s)) \widehat{r}, %otvr moved hat%cdquestion should there be a \hat{r} on the right hand side?
\end{equation}
%\end{widetext}
where % $x=\frac{r}{\sigma}$,
%$v=\frac{C_1}{\sigma}$ is the characteristic velocity,
$v_s^2 \equiv 1/\rho_a \kappa$ and 
the function $H(\xi)$ is defined as 
%in terms of its argument as:
\begin{equation}
H(\xi) = - \frac{e^{\xi^2}}{\xi^2} - \frac{e^{-\xi^2}}{\xi^2} - Ei(\xi^2) -
Ei(-\xi^2).
\end{equation}

\begin{figure}
\includegraphics[width=0.8\textwidth] {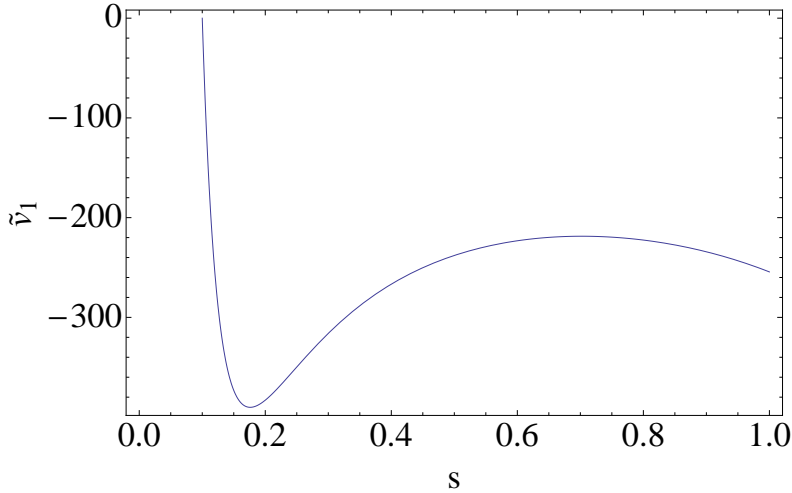}
\caption{The dimensionless first order
correction to the velocity field due to  %otvf
finite compressibility, plotted versus $s$. The quantity
plotted, $\widetilde{v}_1$, is
defined as $v_1/\lambda v$.} %abd edited caption %cdn removed vector sign
\label{fig2}
\end{figure}

In Eq.~(\ref{v1gauss2}) the dimensionless perturbation
parameter $\lambda$ discussed
%in subsection \ref{para}%abdiop
above can be clearly identified as the prefactor appearing in the right side.
%with the prefactor. We now plot in 
Figure \ref{fig2}  shows the magnitude of
the correction to the velocity,
$\vec{v_1}$, in units of 
$\lambda v$ for the case where $C_1=C_2$
as in Eq.~(\ref{v1gauss2}).
This quantity is plotted as a function of the dimensionless
radial coordinate $s$. The value of the cutoff parameter has
been set so that $\delta=0.1$. %abdf
The numbers in the vertical scale seem to
be large. This can easily be seen (see Eq.~(\ref{v1gauss2}))
to arise from the smallness of our choice for the cutoff parameter $\delta$.
We will see that in actual situations, the parameter $\lambda$
is small enough so that the first order velocity  correction
is indeed much smaller than the zeroth order velocity scale. A similar
remark applies to the density correction discussed below.
%\subsection{Circulation}

\begin{figure}
\includegraphics[width=0.8\textwidth] {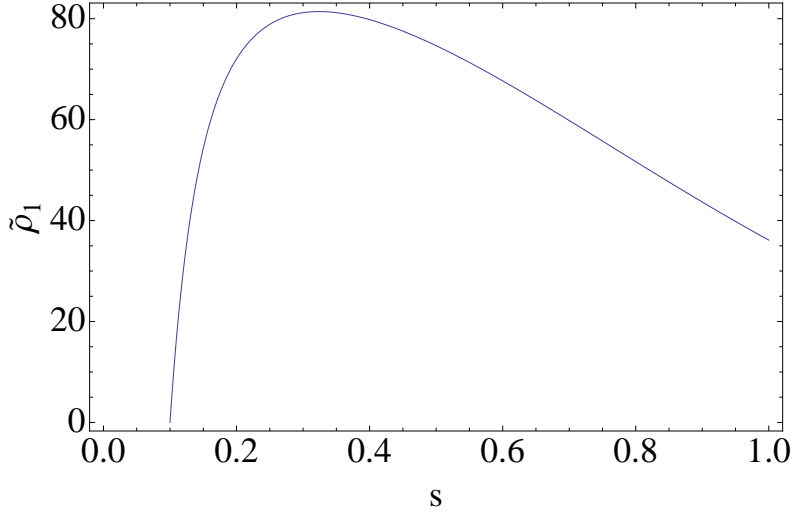}
\caption{Plot of the dimensionless first order 
correction to the density
due to finite
compressibility versus $s$. The quantity plotted, $\widetilde{\rho_1}$, is
defined as $\rho_1/\lambda \rho_a$.}%abd edited caption
\label{fig3} %otv
\end{figure}

We can now study the 
correction to the density profile
($\rho_1$)
and eventually to the physical current %otvf
associated with fluid flow. 
Using Eq.~(\ref{cont1}) 
we find that a solution with azimuthal symmetry
exists and satisfies the first order
inhomogeneous differential equation:
\begin{equation}
\label{rho1gauss}
%\frac{\partial \rho_1}{\partial r} + \frac{2r\rho_1}{\sigma^2} =
%\frac{r\rho_1}{\sigma^2}
-\frac{\partial \rho_1}{\partial r} (v_{0r}) = \rho_0\frac{\partial v_1}{\partial r}
+ \rho_1(\frac{\partial v_{0r}}{\partial r} + \frac{v_{0r}}{r}) + 
\frac{\partial \rho_0}{\partial r}v_{1r}
\end{equation}
It is straightforward to solve this differential
equation  using the method of
integrating factors, and we find
the correction $\rho_1$ to the density %cd
to be:
%\begin{widetext}
\begin{equation}
\label{rho1final}
\frac{\rho_1}{\lambda \rho_a} = \frac{e^{-s^2}}{2}(G_1(\delta) - G_1(s)) %changed e^{s^2} to e^{-s^2} and changed sign for the bracket
\end{equation}
where the function $G_1$ is defined as
%\begin{equation}
\begin{eqnarray}
G_1(\xi) =&& \frac{e^{-\xi^2}}{\xi^2} + \frac{e^{\xi^2}}{\xi^2} -
2\Gamma[0,-\xi^2]-2\log(\xi^2)\nonumber \\
&&- Ei(\xi^2) + Ei(-\xi^2).
%changed sign for gamma function and Ei(\xi^2)
\end{eqnarray}
Here  $\Gamma[a,z]$ represents the incomplete Gamma function.
%defined as:
%\begin{equation}
%\Gamma[a,z] = \int_z^\infty e^{-t} t^{a-1} \mathrm{d}t
%\end{equation}
Figure \ref{fig3} shows a plot of the 
first order correction to the
density. The quantity plotted is the left side
of Eq.~(\ref{rho1final}) as a function of the  dimensionless radial %cd
coordinate $s$ for the same conditions as in Fig.~\ref{fig2}.

\begin{figure*}
%\includegraphics[width=3.0in]{Fig41a.eps}
%\includegraphics[width=2.2in]{Fig41a.eps}
%\includegraphics[width=3in]{vrring2.eps}
%\includegraphics[width=2.2in]{Fig41b.eps}
%\vspace{-2in}
\includegraphics[width=0.8\textwidth]{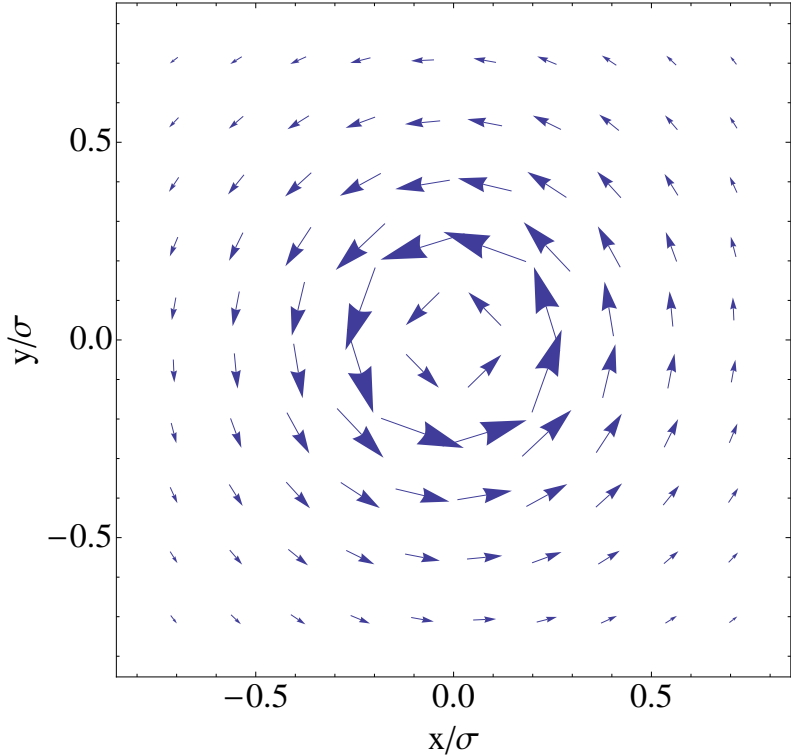}
%\vspace{-0.8in}
\caption{Vector plot of  %otvf cdn2 rewritten caption
the first order correction to the current due to finite compressibility
in the calculation described in subsection~\ref{gaussian}. %otvf 
%Panel A is a vector plot of $\rho_0\vec{v}_1$, %otvf
%panel B  of
%$\rho_1\vec{v}_0$ and panel C of their %otvf
%sum,
%$\vec{j}_1$, due to finite compressibility. 
Lengths are in units
of $\sigma$ and the 
parameters are as 
in Figs.~\ref{fig2} and \ref{fig3}.}
\label{fig4} %otv
\end{figure*}

We can now look  at 
the corresponding current $\vec{j}$  that is
a characteristic physical property 
of the system.
The total current is the sum of zeroth and first
order terms:
%\begin{equation}
%\begin{eqnarray}
$\vec{j}_{total} = \vec{j}_0+\vec{j}_1$, %otvff
%\end{equation}
where $\vec{j}_0=\rho_0 \vec{v}_0$, and %changed o to 0
$\vec{j}_1 = \rho_0 {\vec v}_1 + \rho_1 {\vec v}_0$ 
%\end{eqnarray}
is the first order correction.
%the zeroth order current
%due to compressibility.
%The second order term $\rho_1v_1$ is ignored.
In Figure \ref{fig4}
we present a 
vector plot of $\vec{j}_1$. %cdn2 deleted the first two panels of Fig 4.  
% - $\rho_0\vec{v}_1$, $\rho_1\vec{v}_0$ -  %otvf
%and in the last panel, their sum. 
The conditions, units and parameter values  are as in 
Figs.~\ref{fig2} and \ref{fig3}.
We see that the magnitude of the physical 
current arising from finite compressibility is %otvf
more pronounced closer to the 
central region of the cylinder. 

To check the validity of the perturbative treatment developed here, it is 
necessary to examine the order
of magnitude of the perturbation
parameter $\lambda$ for typical cold atomic systems. We assume that
$C_2$ corresponds to one quantum of circulation. 
Thus we take, as orders of magnitude, 
$\sigma \approx 10^{-5} m$ \cite{gauss1}, $C_2 \approx 10^{-9} m^2/s$ %abdiop2 added the gaussian experiment reference 
and \cite{bec1,lambda, lambda1,speed2} %abdiop1 added more references here
%otviop1H do we need more comment? 
$v_s\approx 10^{-2}
m/s$. This leads to the  estimate  $\lambda \approx 10^{-4}$. 
Thus, for cold atomic systems, while $\lambda$
may be quite small, the corrections are far from negligible since
the quantities plotted in Figs.~ \ref{fig2} and %finaloa
\ref{fig3} %otvf
can be as large as 
several hundreds. Thus, corrections up to the %otvf 
level of $\sim$ 10\%
can easily arise. Hence, we conclude that compressibility effects  in the hydrodynamics of cold atomic systems, %abdiop
%cannot be neglected
as seen from the analytic perturbative method 
used here, %otvr 
cannot be neglected. For superfluid Helium, however, 
the circulation quantum is much larger (a factor of 20
compared to Rb), the speed of sound much larger \cite{speed1} 
and system sizes also larger: thus a similar
estimate yields $\lambda \approx 10^{-14}$ and the corrections
are negligible, as expected.

\subsection{Obstructed Cylinder}

We now consider the second problem, which is perhaps
of clearer physical relevance: we calculate
the compressibility corrections for the
obstructed rotating cylinder geometry studied in Ref.~\cite{usold}.
%whereby the geometry is that of an 
%ideal cylinder - long in the z-direction -
%such that the problem is quasi two dimensional
This geometry is that of a
circular cylinder
of radius $a$  with a thin radial
wall extending from the axis  
to the outer wall of the cylinder. We assume that 
the cylinder is long enough for end effects %cdn2 added a line about end effects
to be negligible. We define an 
angle $\phi$ from the line of obstruction
so that $\phi = 0$ defines the location of the radial %otvf
wall. 
In this case there is no applied force in equilibrium:
thus when the cylinder does not rotate the density
is uniform, at a value which we take as our
unit of density. When the cylinder rotates about its axis with angular 
speed $\Omega$ a velocity field is induced in it.
In the zero compressibility limit, this velocity field
is known\cite{usold}.
We will calculate here the corrections to the 
velocity field and the density profile due to
non-zero compressibility within the perturbative
method described in Sec.~\ref{methods}.  
The field $\vec{v}_0$ 
for the geometry under consideration  %cdn
was obtained in Ref.~\cite{usold} via both scalar and
vector potential methods. It can be expressed in series form:  %cdn
%\begin{widetext}
%otviop\begin{subequations}
\begin{eqnarray} %otviop
\label{vcircle}
%\begin{align}
v_{0r}(r,\phi)=&&\Omega r \sin (2\phi)\nonumber \\ 
&&+\frac{16 \Omega a}{\pi}
\sum_{n\; {\rm odd}}{(\frac{r}{a})}^{n/2-1}\frac{1}{n^2-16}\cos(n\phi/2) %otvr
\end{eqnarray}
\begin{eqnarray}
\label{vcircle2}
v_{0\phi}(r,\phi)=&&\Omega r \cos (2\phi) \nonumber \\ 
&&-\frac{16\Omega a}{\pi}
\sum_{n\; {\rm odd}}{(\frac{r}{a})}^{n/2-1}\frac{1}{n^2-16}\sin (n\phi/2). %otvr
%\end{align}
\end{eqnarray}
%otviop\end{subequations}
%\end{widetext}
%where $\rho \equiv r/a$.
%abd2 replaced \rho by r/a in equation and removed definition of \rho
To obtain the 
correction to the 
zeroth order velocity field ($\vec{v}_1$)
due to finite compressibility 
we solve Eq.~(\ref{ns1u}) by the Green function %otvf
method. We introduce a scalar velocity potential
$V_1$ so that $\vec{v}_1({\bf r})=\nabla V_1({\bf r})$, %abd2 removed \vec from nabla
and calculate the Green function
associated with the operator $\nabla^2$
(see  
Eq.~(\ref{potential})) for appropriate boundary
conditions. We recall\cite{usold} that the
boundary condition on the total velocity field, 
%with a Neumann boundary condition on
%$V_1$: %(recalling that
%${\bf v}({\bf r}) = \nabla V({\bf r})$):
%\begin{equation}
${\bf v}({\bf r})_\perp=({\bf \Omega} \times {\bf r})_\perp$,
%\label{bc}
%\end{equation}
where ${\bf r}$ is a vector
from the center to a point on the
boundary, and $\perp$ denotes the component normal to the 
boundary, %${\bf \Omega}$ is the angular %cdn2 added a line to define \perp
%velocity of the container holding the 
%fluid and ${\bf v}$ being the velocity field 
%of the fluid.
is already
satisfied by the zeroth order velocity in Eq.~(\ref{vcircle}). % 
Hence, $V_1({\bf r})$ satisfies zero Neumann boundary conditions %otvf %cdn2 check this statement
at the cylinder surface. %checked ({\bf CHECK THIS STATEMENT}) 
The Green function 
in this case is then found by standard procedures\cite{jackson}
with the result:
%otviop \begin{widetext}
\begin{equation}
G(r,\phi;r^\prime,\phi^\prime)= -\frac{1}{\pi} \sum_{n=2}^\infty \frac{1}{n}
 r_<^{n/2}\left(\frac{1}{r_>^{n/2}} + \frac{r_>^{n/2}}
{a^{n}} \right) \cos(n\phi/2) \cos(n\phi'/2), %abd2 added prime(') to \phi
\label{gf3}
\end{equation}
%otviop \end{widetext}
where $r_>$ ($r_<$) is the larger (smaller)
one of the two radial coordinates
$r$ and $r^\prime$. 
One then gets an expression for 
the 
velocity potential
in the form of 
Eq.~(\ref{po2}), with $G(\vec{r},\vec{r}^\prime)$ given by Eq.~(\ref{gf3})
and the $\vec{v}_0.(\vec{v}_0.\nabla)\vec{v}_0$ evaluated
from Eqns.~(\ref{vcircle}) and (\ref{vcircle2}). Taking then the 
gradient, the correction to the velocity
field is calculated. 
%otvr new sentences below
In our subsequent calculations, we introduce a
radial cutoff, which we take to be $0.1 a$, %otvf
to exclude %cdn changed wording to avoid repeat of exclude
the small $r$ region where\cite{usold} the zeroth order velocity has a
weak square root singularity.

In principle, this procedure involves  no
advanced mathematical steps. However, one can readily see that
it is very lengthy and intricate. Since the right side of Eq.~(\ref{po2})
is cubic in $\vec{v}_0$, the components of which are in the form
of a series, and the Green function Eq.~(\ref{gf3})  involves an additional
sum, the expression has the form of a quadruple sum, plus integrals
over the angle $\phi^\prime$ and the radial coordinate 
$r^\prime$. %otvff added some of it back %cdn2 deleted text describing technical details
This is done analytically: the angular integrals are performed first and
lead to Kronecker deltas that reduce the number of sums. The radial
integrals are done next and finally, the remaining (convergent) summations
are evaluated.
%The most efficient
%way to proceed is to do the angular integral first. This leads 
%in general to
%combinations of Kronecker delta functions which reduce the number
%of sums to be performed, although at the price of a large increase
%in the number of terms to be computed. The radial integrals
%can then be performed analytically leading to combinations of
%powers of $r$, and the evaluation
%of the gradient of $V_1$ poses no
%difficulties. The multiple summations mentioned still remain to
%be done, but they can be shown to be convergent (once the terms are
%properly reorganized) for all values of $r$ except
%near $r=0$ excluded by the cutoff, %otvf
%where the result of course diverges in a small region
%where the assumptions\cite{usold} are not valid. %finaloa
The resulting expression, however, %otvff
is much too involved to be written here, and it would be truly very
difficult even to keep track of all the terms %otvff
without the help of a symbolic package
(we have used Mathematica). 
%The same package is then used to evaluate
%the convergent sums, as a function of $r$ and $\phi$.  %cdn2 deleted text describing technical details
The results can then be plotted and the plots are much more illuminating
than the lengthy expressions.

One can however see from the basic
structure of the equations that 
overall $\vec{v}_1$ is proportional to the basic
velocity scale of the system, which is $\Omega a$, times
a factor of $\lambda = ({\Omega a}/{v_s})^2$ where %otvr %abd2 modified a factor to lamba of order unity
$v_s^2=1/(\kappa\rho_0)$.  %finaloa  
It is rather obvious
also from the equations 
that $\vec{v}_1$ has both radial and azimuthal components.
Results for these components are
shown in the next two figures. There, the 
dimensionless quantity plotted
is a component of $v_1$, divided by $\lambda \Omega a$.
These are shown as functions of angle and dimensionless
radial distance $r/a$. 

\begin{figure}
\includegraphics[width=0.8\textwidth] {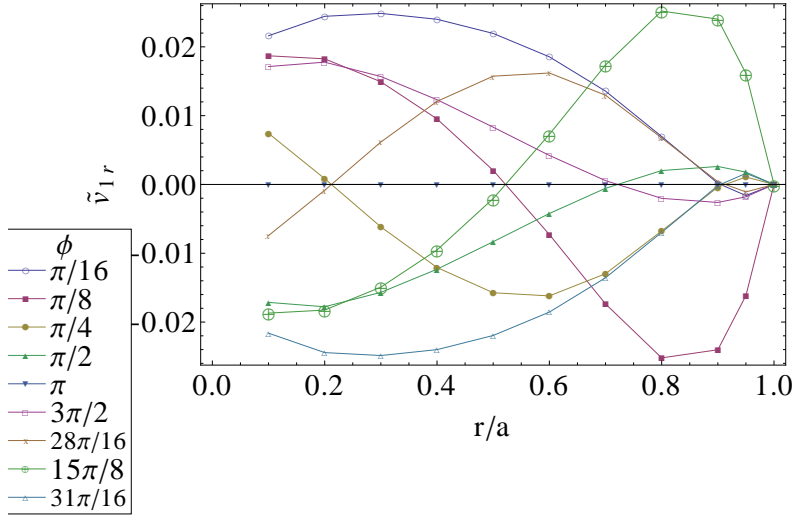}
%abdf caption redone
\caption{Plot of the  
radial component of the 
correction to the velocity field $\vec{v}_1$ for an obstructed
cylinder. The dimensionless quantity shown is  
${\widetilde{v}}_{1r} \equiv {v}_{1r}/(\lambda v)$, %cdf put back \vec
%because only a vector can have a radial component.
plotted versus $r/a$
at different angles $\phi$.} %otvf
\label{fig6}
\end{figure}

\begin{figure}
\includegraphics[width=0.8\textwidth] {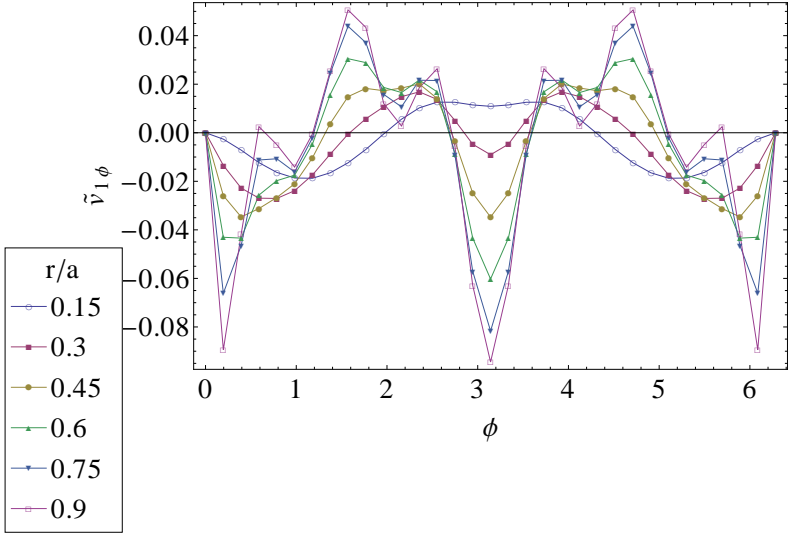}
%abdf redone
\caption{Plot of the 
azimuthal component of the
correction to the velocity field $\vec{v}_1$. The dimensionless quantity shown is
${\widetilde{v}}_{1\phi} \equiv {v}_{1\phi}/(\lambda v)$, plotted  %cdf put back \vec
%because only a vector can have an azimuthal component.
versus $\phi$
at different values of $r/a$.}
\label{fig7}
\end{figure}

In  
Figure \ref{fig6} we present a plot 
of the radial component ($v_{1r}$) 
in the units described above, at several fixed
values of the angle $\phi$.
One can see that 
the radial correction to the velocity %otvr
field goes to zero at $r/a=1$  satisfying %otvf
the radial boundary condition discussed above. %abd2 added a revised description of v_1r plot
The magnitude of the correction to the radial velocity ($v_{1r}$)  %otvr
peaks at different values of $r/a$ depending on 
how far one is from the line of obstruction along the azimuthal direction.
At $\phi = {\pi}/{8}$, %otvr 
the peak in the magnitude of $v_{1r}$ occurs %otvr %cdn
closer to $r/a=1$ than  at $\phi = {\pi}/{2}$ where it %otvf 
occurs closer %otvr %cdn
to the lower cut-off. 
At $\phi=\pi$,  $v_{1r}$ is identically zero %otvf
owing to the symmetry of the problem over an angle of $2\pi$.
%abdr add discussion of the interesting result mentioned in intro here? 
The corresponding azimuthal component
($v_{1\phi}$)
is presented in Figure \ref{fig7}.
The quantity $v_{1\phi}$ is now plotted in dimensionless
units  at several values 
of the dimensionless distance $r/a$. It can be seen that
%otvr rewritten below
oscillations, in which one can discern traces corresponding to the 
$n=3$ and $n=5$ terms of the expression for $v_{0\phi}$ in Eq.~(\ref{vcircle}), %changed v_{0,\phi} to v_{0\phi} 
%abd2 added plus n=5 to explain the revised figure %otvr more rewrite
are present in %otvf %cdf changed "visible" to "present" 
%because the sentence starts with "it can be seen"
the azimuthal correction to the velocity. This arises from the
cubic (in $v_0$) nature of the perturbation.
Again, the boundary conditions at 
$\phi=0,2\pi$ are satisfied with the velocity
correction being zero at these values of $\phi$. 

Having calculated the 
correction to the velocity field %otvr
due to finite compressibility, we 
can next study the correction to the density %otvf
profile
($\rho_1$). This turns out to be computationally
much simpler. %abd2 changed the explanation of \rho_1 calculation below
Starting with Eq.~(\ref{rho1}), a line integral over
$d\vec{r}$ enables us to calculate $\rho_1$.
Since the right side of Eq.~(\ref{rho1}) can be expressed as the %otvr
gradient of a scalar quantity, doing an integral
over $\vec{r}$,
\begin{equation}
\int \nabla\rho_1\cdot d\vec{r}= -\int
\rho_0^2\kappa(\vec{v}_0\cdot\nabla)\vec{v}_0\cdot d\vec{r} %otvr sign
\end{equation}
enables us to calculate $\rho_1$ independent of the 
path chosen within the obstructed cylinder. 
Using this property of $\nabla\rho_1$, we calculate 
$\rho_1(r,\phi)$ from a line integral over 
two different paths and confirm
that our result is indeed path independent. 
We determine the arbitrary constant associated with the 
integration by enforcing 
the condition that the total integral of $\rho_1$ over the
relevant region be zero i.e.
%\begin{equation}
$\int \rho_1rdrd\phi = 0$. %otv no set off and some rewrite
%\end{equation}
This condition makes sense physically since the constraint imposed by the
container implies that the total mass change due to the %otvr 
compressibility correction must be zero.  %otvr
\begin{figure}
\includegraphics[width=0.8\textwidth] {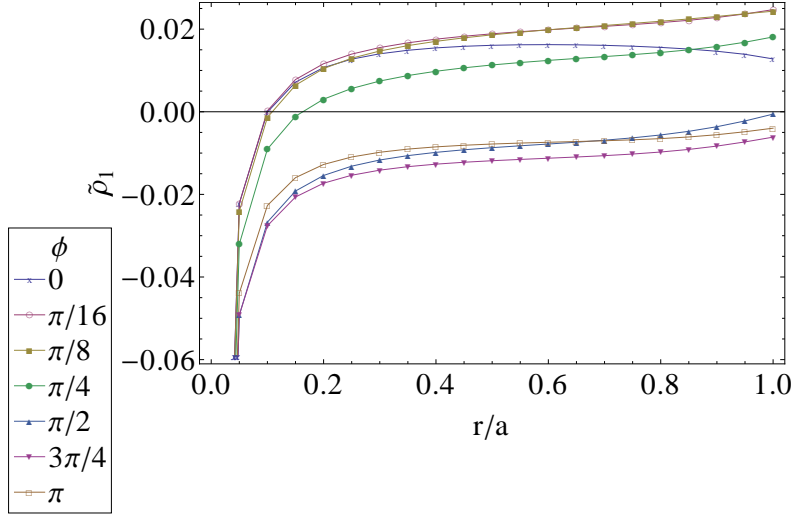}
\caption{Plot of the 
radial dependence of the dimensionless correction to the density -
$\widetilde{\rho}_1 \equiv {\rho_1}/{(\lambda \rho_0)}$ -
due to finite compressibility
at different values of $\phi$.}
\label{fig9} %otv
\end{figure}
This gives us the full function $\rho_1(r,\phi)$. %otvf no new para 
Because of the nonlinearities in $v_0$ present in the 
equations, the results, although formally
analytic in terms of convergent double series, are again quite intricate 
and will not be written down explicitly here.
Instead, as  before, results are plotted in the next two
figures. The quantity plotted is the dimensionless
$(\rho_1/(\rho_0 \lambda))$
and we plot it at different values 
of $\phi$ in Figure \ref{fig9}.
One can see that that radial 
dependence of the density correction is more
prominent closer to the line of 
obstruction within the cylinder, which makes sense
physically. Also, it is
positive or close to zero 
near the outer boundary, and negative at smaller
values of $r/a$. %abd2 changed description of figure 7
%abdiop1 added more detailed description of result
%otviop1H edited
This seems to
be a direct consequence of the fact that the higher 
the velocity of
fluid flow, the lower is the condensate density \cite{raman}.
Physically, the compressibility correction depletes the density
at smaller values of $r/a$  and, conversely, increases it at 
large values.
%abdiop1 end addition here
Similarly, the azimuthal dependence
of $\rho_1(r,\phi)$, 
is plotted  at different values of 
$r/a$
in Figure \ref{fig8}.
\begin{figure}
\includegraphics[width=0.8\textwidth] {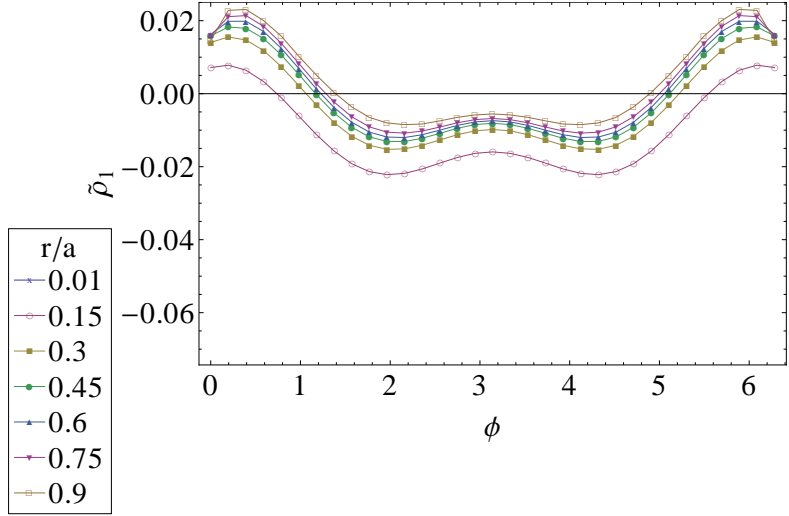}
\caption{Plot of  the 
azimuthal dependence of the dimensionless correction to the density - 
$\widetilde{\rho}_1 = {\rho_1}/{(\lambda \rho_0)}$ -
due to finite compressibility 
at different values of ${r}/{a}.$}
\label{fig8} %otv
\end{figure}
The angular dependence of these results reflects again
the nature of the perturbation.

We proceed now to analyze the order of magnitude of
these corrections in typical cold atomic systems. For this
problem (as opposed to the gaussian profile situation)
the values of the dimensionless quantities plotted do not exceed
unity, so that the corrections
are now smaller and the perturbation theory
is valid for larger values of $\lambda$. 
%abdiop1 added following sentence 
%otviop1I
Our results show that compressibility corrections 
for fixed $\Omega$ %cdiop1
are smaller for more confined geometries. 
%abdiop addition ends here
In this case, it is best to phrase
the issue in terms of the validity of the
theory (or the need for it) in the upper part of the relevant range
of $\Omega$. This range of $\Omega$ values is limited
by the requirement that the circulation created
by the zeroth order velocity does not exceed a flux
quantum. The velocities involved are roughly\cite{usold}
of order $10^{-1}\Omega a$. Demanding then
that this speed, times $a$, be of order of
one flux quantum and using the
order of magnitude 
values discussed
previously, we find that the corresponding $\lambda$
for cold atomic 
systems may reach values up to $10^{-2}$. Thus, we find that for these
systems finite
compressibility corrections are not always negligible but on the other
hand they are, at 
least for samples that are not too small, amenable to our perturbation
approach.  For superfluid helium we
find, as expected, that $\lambda$ is much smaller and the 
incompressible limit calculations are perfectly adequate.

\section{Conclusions}
\label{conclusion}
In this paper, we have studied, by means of an analytic method, %abdiop
the hydrodynamics of compressible
superfluids in confined geometries. We have shown that for 
practical cases of interest in cold atomic systems 
confined to complex geometries the corrections to %abdiop
the zero compressibility results are not negligible, but they can
in many realistic cases %otviop1AG
be treated in a perturbative manner with the relevant 
dimensionless parameter being
the square of the ratio of the typical speed to the  sound speed 
(the Mach number).
This method may be used in very general situations.
We
have illustrated the procedure by working out two examples. In the first,
confining forces that need not be specified constrict the 
%equilibrium (non-moving) %cdq since v_0 is not zero, can we call the fluid non-moving? 
fluid to having a Gaussian density distribution
at zeroth order. %finaloa
In the second, we have considered
the case of a rotating obstructed
cylinder filled with superfluid, with the density being
uniform when the cylinder is at rest. The zeroth order (incompressible
limit) solution to the problem is available\cite{usold} and the perturbative
method is used to evaluate the corrections, again essentially in an
analytic way. The general usefulness of the method is therefore illustrated
by these examples. The general nature of the 
perturbative method applied to superfluids in confined %abdiop
geometries makes it
useful for describing the results 
of relevant experiments on cold atomic systems, %abdiop 
some of which have been mentioned in the Introduction.
%otviop1AG
One of the main advantages of our procedure 
is that, since the resulting
equations are linear, it is very amenable to
analytic solution. %otviop1AG
We expect that this %cdnew new text
method will complement numerical calculations 
based on more microscopic descriptions such as the GP equation.  
%otvff reworded last sentence[{\bf I DON'T UNDERSTAND THE LAST SENTENCE}]
Starting directly from the hydrodynamic equations is, in the
appropriate limit, a good alternative to using the full GP
equations, for situations where a hydrodynamic scale 
flow is imposed on the system. %otviop1AG
%and also serve as an alternative to analytic studies of cold atomic systems using %abdiop
%GP equation in the hydrodynamic regime.
%abd2 replaced AMO by cold atomic

%Corrections negligible for He. Old calculation correct.

%Perturbation theory appropriate for AMO realistically.

\section{Acknowledgment}
This research was supported in part by IUSSTF grant 94-2010.

\end{document}